\documentclass{article}
\usepackage{amssymb}
\usepackage{amsfonts}
\usepackage{amsmath}
\usepackage{graphicx}
\begin{document}

\title{Impact of edge-removal on the centrality betweenness of the best spreaders}

\author{N. N. Chung$^1$, L. Y. Chew$^2$, J. Zhou$^1$ and C. H. Lai$^{3,4}$ \\
\\
$^{1}$ Temasek Laboratories,\\
 National University of Singapore, \\
 Singapore 117508 \\
$^{2}$ School of Physical \& Mathematical Sciences, \\
Nanyang Technological University,\\
 21 Nanyang Links, Singapore 637371 \\
$^{3}$Beijing-Hong Kong-Singapore Joint Centre for \\
Nonlinear and Complex Systems (Singapore), \\
National University of Singapore, \\
Kent Ridge 119260, Singapore \\
$^{4}$ Department of Physics, \\
National University of Singapore,\\
 Singapore 117542}
\maketitle

\begin{abstract}
The control of epidemic spreading is essential to avoid potential fatal consequences and also, to lessen unforeseen socio-economic impact. The need for effective control is exemplified during the severe acute respiratory syndrome (SARS) in 2003, which has inflicted near to a thousand deaths as well as bankruptcies of airlines and related businesses. In this article, we examine the efficacy of control strategies on the propagation of infectious diseases based on removing connections within real world airline network with the associated economic and social costs taken into account through defining appropriate quantitative measures. We uncover the surprising results that removing less busy connections can be far more effective in hindering the spread of the disease than removing the more popular connections. Since disconnecting the less popular routes tend to incur less socio-economic cost, our finding suggests the possibility of trading minimal reduction in connectivity of an important hub with efficiencies in epidemic control. In particular, we demonstrate the performance of various local epidemic control strategies, and show how our approach can predict their cost effectiveness through the spreading control characteristics.
\end{abstract}

\newpage

It had been shown that mathematical models that take full consideration of the complexity of aviation networks can be used to obtain a detailed forecast of the epidemics in a globalized world. In particular, simulation results of the SARS outbreak are shown to be in good agreement with the reported cases \cite{Hufnagel04}. There are also empirical evidences of the direct influence of airline travel on dissemination of the seasonal influenza in the United States \cite{Brownstein06}. Hence, if we can change the topological structure of the transportation networks, we are able to control the epidemics. In fact, finding an efficient way to slow down the propagation of infectious diseases within a society has always been an important subject in network sciences \cite{Hufnagel04,Brownstein06,Marcelino09,Holme02,Pastor02,Kitsak10,Spreading_lit,Crepey07}. In these studies, strategies like removing the highly connected nodes \cite{Holme02,Pastor02} or edges with high centrality betweenness \cite{Marcelino09,Hufnagel04} are proposed to control the spreading of diseases. Although both methods are shown to be effective in hindering the propagation of diseases, both introduce unavoidably enormous costs to the society. As an example, the seasonal influenza hits the highest number of infected cases each year during the winter and displays typically a seasonal cycle \cite{Brownstein06}.  The beginning of the major influenza activity can be foreseen and even tracked geographically from time to time, yet it is responsible for the death of 3000 to 49,000 people in the United States each year \cite{CDC}. In fact, Cr\'{e}pey and Barth\'{e}lemy \cite{Crepey07} have studied 30 years of data on the seasonal influenza in the United States and found that sensible modeling of the epidemics may only need to include the air transportation data. Why are epidemics control strategies like removal of nodes or edges with the strongest connections not used before the approaching of the influenza? As we all know, this is not practical as it will cost billions of dollars to the affected airports. Thus, it is important to have a comprehensive way to analyze the trade-off between the effectiveness and the cost to pay before a change in the network topological structure is employed for epidemics control.

In this article, we study the impact of edge-removal in terms of both the cost and the effectiveness. Our study focus on the best spreaders in a network since they can play more crucial roles than the others during an epidemic outbreak. As suggested by Kitsak et. al. \cite{Kitsak10}, the best spreaders are influential nodes located within the network's core which can be identified using the k-shell decomposition analysis \cite{k-shell}. Among the many connections attached to the best spreader, our aim is to uncover the role of each edge by analyzing the effectiveness in slowing down the spreading and the resulting cost to be paid upon removal of the edge through defining appropriate quantitative measures. We consider the spreading control effectiveness of the topological changes as measured by the decrease in the extreme eigenvalue ($\lambda_m$) of the network adjacency matrix whereas the cost of an edge-removal is quantified by the decrease in the centrality of the influential spreader.

Explicitly, upon removal of an edge $ij$ attached to a best spreader $i$, we quantify the relative effectiveness ($E_{ij}$) in increasing the epidemic threshold by the decrease in the extreme eigenvalue ($\Delta \lambda_{ij}$), normalized by the decrease in $\lambda_m$ upon removal of node $i$ ($\Delta \lambda_i$), i.e.
\begin{equation}
E_{ij}= \frac{\Delta \lambda_{ij}}{\Delta \lambda_i} = \frac{I_{ij}}{I_i}\,.
\end{equation}
Note that $I_{ij}$ and $I_i$ denote the dynamical importance \cite{Restrepo06} of edge $ij$ and node $i$ respectively. Since the epidemic threshold is shown to be inversely proportional to the maximum eigenvalue \cite{EpidemicThreshold}, $E_{ij}$ captures how changes in the topological structure increases the difficulty for the epidemic outbreak to take place.  On the other hand, the cost ($C_{ij}$) of an edge-removal to the influential spreader is quantified by the normalized decrease in the node's betweenness:
\begin{equation}
C_{ij}= \frac{\Delta B_{ij}}{B_i} \,.
\end{equation}
Centrality betweenness of a node measures the node's relative importance in the network by counting the number of shortest paths that cross through the node \cite{Newman01}. Isolation of an influential node is equivalent to removing all the links attached to it, and the betweenness of the node thus drops to zero. Meanwhile, removal of edges introduces partial inactivation of the connection between the node and its nearest neighbors such that functionality of the node remains with reduced betweenness.

\begin{figure}
\begin{center}
\includegraphics[scale=0.6]{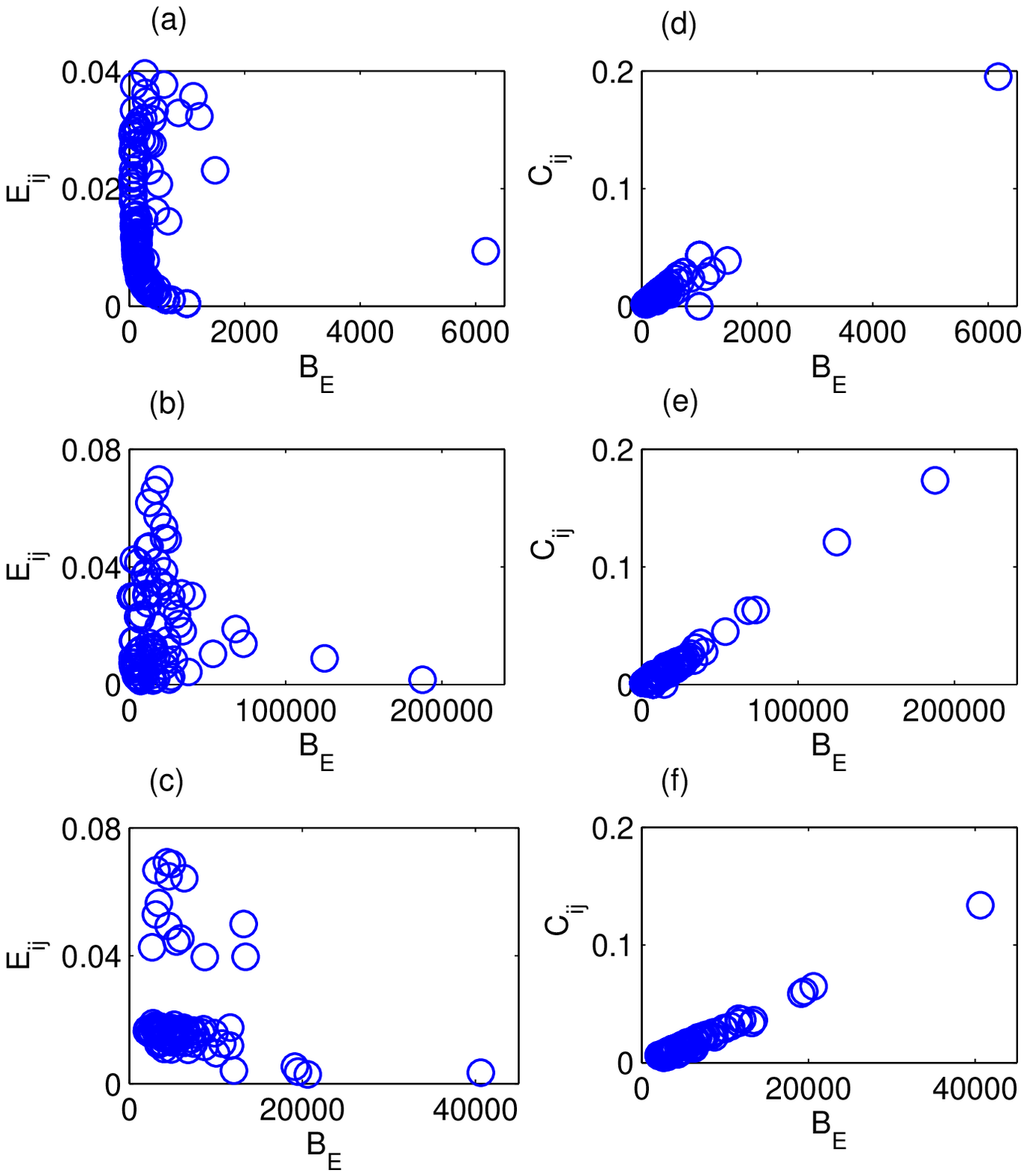}
\end{center}
\caption{Dependence of $E_{ij}$ and $C_{ij}$ on $B_E$ of the removed link for (i) the US air transportation ((a) and (d)), (ii) the collaboration((b) and (e)) and (iii) the Gnutella ((c) and (f)) networks.} \label{fig1}
\end{figure}

We study the impact of edge-removal on three real-world complex networks from different fields, i.e.: (1) the US air transportation network \cite{Colizza07}, (2) the collaboration network in computational geometry \cite{Jones02} and (3) the Gnutella peer-to-peer internet network \cite{Gnutella}. The airline network is an undirected network obtained by considering $500$ airports in US that have the largest amount of traffic. This kind of network is important for the spreading of infectious diseases such as the influenza and SARS. In addition, it illustrates a good example where removal of high degree nodes and high betweenness edges comes with huge economic and social cost to the affected airports. The collaboration network describes the collaboration relation between $6158$ authors. By considering this network, the spreading of rumors or ideas, and the control of their spread through edge-removal, can be studied from the context of a social network. Lastly, the Gnutella network, which is constructed from a sequence of snapshots of the Gnutella peer-to-peer file sharing network, is included for the study of computer virus spreading and its control. For each network, we choose an influential spreader for edge-removal analysis. Then, for all the edges attached to the influential spreaders, we find the edge-betweenness ($B_E$) of each edge, i.e. the number of shortest paths that cross through the link and compute $E_{ij}$ as well as $C_{ij}$. Figures \ref{fig1} (a) - (c) show the spreading control effectiveness of each edge-removal with the corresponding edge-betweenness. In contrast to common belief, removal of the link with the highest value of $B_E$ does not give rise to the largest reduction in $\lambda_m$. Instead, removals of many of the other links attached to the best spreader which have much lower value of $B_E$ are more effective in lowering the value of $\lambda_m$. Nonetheless, there is no unique relation between $E_{ij}$ and $B_E$ since removal of edges with similar $B_E$ can cause very different amounts of reduction in $\lambda_m$ .

\begin{figure}
\begin{center}
\includegraphics[scale=0.5]{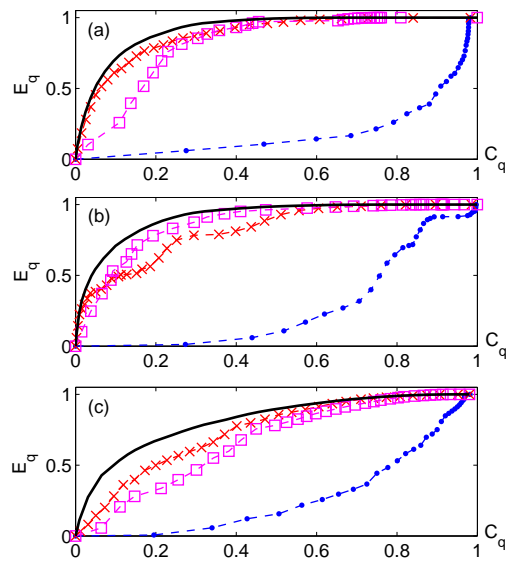}
\end{center}
\caption{Spreading control characteristics for the selected influential spreaders in (a) the US air transportation, (b) the collaboration and (c) the Gnutella networks. Note that SCC are plotted as dashed curves with dots, dashed curves with crosses, dashed curves with squares and solid curves for the first, second, third and fourth strategies respectively. } \label{fig2}
\end{figure}

Next, we consider two spreading control strategies, namely, removing the edges attached to the best spreader one by one following the decreasing and increasing order of $B_E$. Suppose that the maximum eigenvalue drops by an amount of $\Delta_q \lambda$ and the centrality betweenness of the best spreader drops by $\Delta_q B$ after removing $q$ links (following a specific order according to the adopted strategy) from node $i$, we then have
$
E_q=\Delta_q \lambda/\Delta \lambda_i \,,
$
and
$
C_q=\Delta_q B/B_i \,.
$
As more edges are removed, the resulting spreading control effectiveness, $E_q$ increases, at the same time, the resulting cost $C_q$ also increases. We propose to characterize this by an $E_q-C_q$ plot which we term the spreading control characteristics (SCC). As shown in Fig. \ref{fig2}, the second strategy, i.e. removing the edges following the increasing order of $B_E$ is more efficient. It offers higher effectiveness with smaller cost. In fact, the performance of both strategies in reducing $\lambda_m$ are comparable when equal number of links are removed. Nevertheless, as shown in Figs. \ref{fig1} (d) - (f), impact on the centrality of the best spreader is proportional to $B_E$ of the removed link. This explains why it is possible to achieve similar epidemic control effectiveness with smaller cost by removing links with smaller $B_E$.

\begin{figure}
\begin{center}
\includegraphics[scale=0.5]{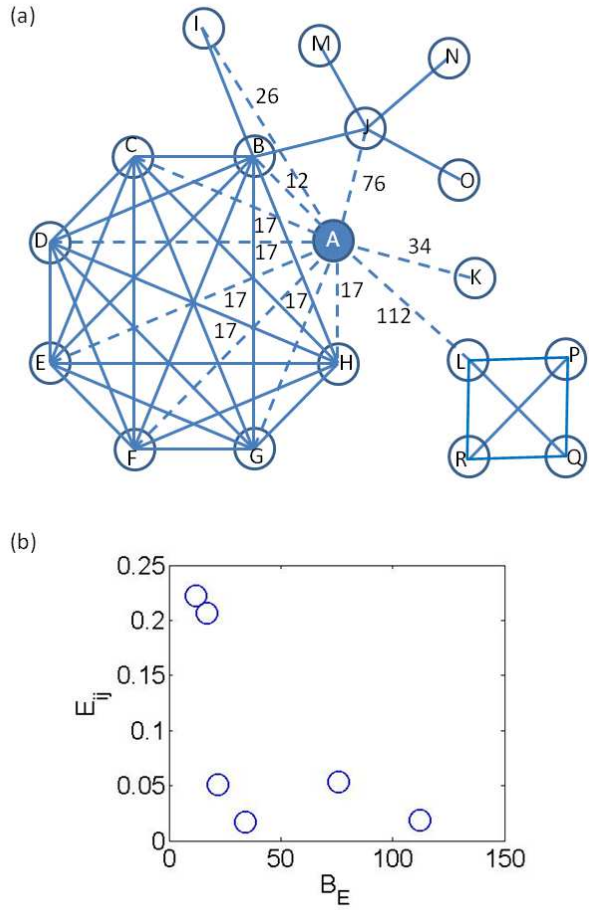}
\end{center}
\caption{(a) Illustration of a simple network which consists of $18$ nodes and $71$ edges. Note that the eleven edges linked to node $A$ are shown as dashed lines and the numbers assigned to them denote edge-betweenness of the links. The other edges are shown as solid lines. (b) Dependence of $E_{ij}$ on $B_E$ of the removed link for the network shown in figure (a).} \label{fig3}
\end{figure}

Edge-betweenness measures the relative importance of a link in the network by counting the number of shortest paths that go through the link. Our study shows however that removal of the popular links does not necessary help much in slowing down the spreading. Why is this so? Let us look at a simple graph in Fig. \ref{fig3} (a). In this network, diseases can spread efficiently within the fully connected subgraph $ABCDEFGH$ as there exist many pathways through which viruses can infect nodes in the subnetwork. For an airline network, this subgraph corresponds to major airports that have direct connections to each other. Passengers can travel freely and easily between any two major airports through one flight and this enhances the spreading of diseases within these major airports. We note that nodes in this subnetwork have higher $k_s$ values and according to Kitsak et. al \cite{Kitsak10}, they are the best spreaders. If one of these nodes is infected, since this infected node is connected to many other influential spreaders, there is a very high chance that another influential spreader is infected. Viruses then propagate quickly out from this subgraph.
In contrast, if the infected node has a low $k_s$ value, diseases propagate less efficiently since neighbors of this infected node are less influential in the process of spreading.

Next, we investigate $B_E$ of the links. Among the edges attached to node $A$, link $AL$ has the largest $B_E$ as it is the only connection to the cluster $LPRQ$. Any shortest path to this cluster has to cross through this link. Thus, Link $AL$ is ``important''. Meanwhile, although link $AJ$ is connected to an equal-size cluster, it is not the only connection to cluster $JMNO$, $B_E$ of link $AJ$ is hence smaller than that of link $AL$. Lastly, betweenness of links connecting node $A$ to nodes $B, C, D, E, F, G$ and $H$ are the smallest as the shortest paths distribute among the many links attached to them. Nevertheless, removal of these links results in the largest decrease in $\lambda_m$. In this case, removal of the edge with the lowest $B_E$ yields the highest effectiveness in spreading control. If the flights between airports $A$ and $B$ are canceled, travelers going to the other major airports will still need to go through airport $A$, hence, centrality of the airport does not decrease much but the probability to have more influential spreaders get infected decreases and epidemic spreading becomes more controllable. On the other hand, removing edges $AJ$ or $AL$ is less effective in controlling epidemic spreading since they are attached to the less influential nodes $JMNO$ or $LPRQ$ which have lower $k_s$ values. Futhermore, removing them impedes passengers from the small clusters to travel through airport $A$ to the other majors airports, which leads to significant drop in the centrality of airport $A$. In Fig. \ref{fig3} (b), we show the $E_{ij}$-$B_E$ plot for this simple graph which is similar to those for real-world networks.

To increase $E_q$, we removes the edges following the decreasing order of $k_s$ of the nearest neighbors. SCC of this strategy are shown as curves with squares in Fig. \ref{fig2}. Although this third strategy is more effective in spreading control compared to strategy 2, the cost to pay is not minimized. Therefore, a fourth strategy is introduced to optimize the result. For this, we define a gain in spreading control for each edge attached to the best spreader as: $G_{ij}=E_{ij}/C_{ij}$.
We rank edges with finite $G_{ij}$ following the decreasing order of their gain. Edges are then removed successively according to their ranking. As shown in Fig. \ref{fig2}, SCC for this strategy is the best among the four strategies considered. The performance of this strategy differs however in the three networks. It is better in the airline network (AUC=0.9146) and the collaboration network (AUC=0.9138) but not so good in the Gnutella network (AUC=0.8184). Here, AUC is the area under the SCC curve which we used to quantify how good a SCC curve is. In the airline network, since $k_s$ value of the best spreader is larger, there are more links with small $C_{ij}$ but large $E_{ij}$ attached to it, they are those located at the top left corner in Fig. \ref{fig1} (a). Since the removal of these links yields larger spreading control effectiveness with less cost, the area under the SCC curve is larger for the airline network.

We have also examined the epidemic control effectiveness of these edge-removal strategies in terms of the delay in the spreading time with the best spreaders as starting nodes. For this, the standard Susceptible-Infected (SI) model with infectious probability $\beta=0.01$ is used. Upon the removal of $q$ number of links, the spreading time until half of the network are infected increases by an amount of $\Delta_q T$ and the spreading control effectiveness is measured by $F_q=\Delta_q T/T_0$. Here, $T_0$ denotes the original spreading time when no link is being removed from the network. The SCC curves in terms of $F_q$ are shown in Fig. 4. As shown in Fig. 4, strategy 2 is more cost-effective than strategy 1 not only in increasing the epidemic threshold but also in increasing the spreading time. In addition, similar results are also obtained for strategies 3 and 4 in terms of the delay in the spreading time.

In summary, due to the complexity in the topological structure of real-world networks, the effectiveness of edge-removal on epidemic control depends nontrivially on the betweenness of the removed edges. A comprehensive analysis is thus necessary before an edge-removal strategy is employed for epidemic control. For this, we suggest to compare spreading control characteristics of different epidemic control strategies for the network concerned. In our examples, we show that SCC can be different for distinct networks or edge-removal strategies. In general, removing edges with the highest gain yields better performance in maximizing the effectiveness of epidemic control with minimal cost. \\
\\

\noindent
{\bf \large{Acknowledgement} } \\

This work is supported by the Defense Science and Technology Agency of Singapore under project agreement of POD0613356.

\newpage

\begin{figure}
\begin{center}
\includegraphics[scale=0.5]{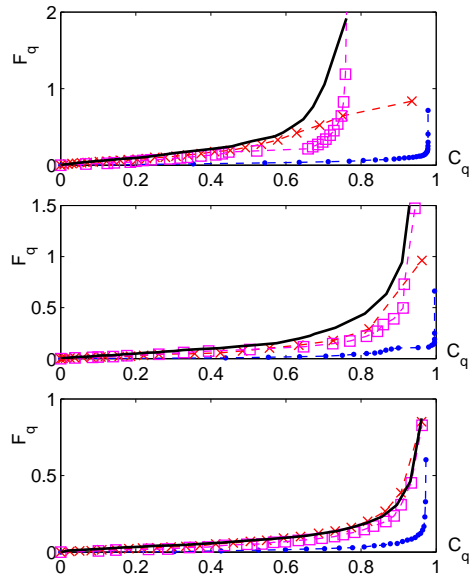}
\end{center}
\caption{SCC in terms of $F_q$ for the selected influential spreaders in (a) the US air transportation, (b) the collaboration and (c) the Gnutella networks. The SCC curves are plotted as dashed curves with dots, dashed curves with crosses, dashed curves with squares and solid curves for the first, second, third and fourth strategies respectively. Note that the results are based on 2,000 different realizations for each data point. } \label{fig4}
\end{figure}

\end{document}